\documentclass[notitlepage,twocolumn,prl,nobalancelastpage,amssymb,superscriptaddress,showpacs,longbibliography]{revtex4-1}
\pdfoutput=1

\usepackage{Symbols}
\usepackage{Shortcuts}
\usepackage{Text}

\usepackage{stmaryrd}
\usepackage{epsfig}
\usepackage{subfigure}
\usepackage[colorlinks=true,citecolor=blue,linkcolor=blue]{hyperref}
\usepackage{verbatim}
\usepackage{xcolor}

\def \vc{{\boldsymbol{c}}}
\def \vy{{\boldsymbol{y}}}
\def \vu{{\boldsymbol{u}}}

\def \vk{{\boldsymbol{k}}}
\def \vr{{\boldsymbol{r}}}

\def \vq{{\boldsymbol{q}}}

\def \vx{{\boldsymbol{x}}}

\def \vsi{{\boldsymbol{\s}}}

\begin{document}
\title{Generating coherent phonon waves in narrow-band materials: a twisted bilayer graphene phaser}

 \author{Iliya Esin}
 \affiliation{Department of Physics and Institute for Quantum Information and Matter, California Institute of Technology, Pasadena, California 91125, USA}
 \author{Ilya Esterlis}
 \affiliation{Department of Physics, Harvard University, Cambridge MA 02138, USA}
 \author{Eugene Demler}
 \affiliation{Institute for Theoretical Physics, ETH Zurich, 8093 Zurich, Switzerland}
 \author{Gil Refael}
 \affiliation{Department of Physics and Institute for Quantum Information and Matter, California Institute of Technology, Pasadena, California 91125, USA}
 \date{\today}

\begin{abstract}
Twisted bilayer graphene (TBG) exhibits extremely low Fermi velocities for electrons, with the speed of sound surpassing the Fermi velocity. This regime enables the use of TBG for amplifying vibrational waves of the lattice through stimulated emission, following the same principles of operation of free-electron lasers. Our work proposes a lasing mechanism relying on the slow-electron bands to produce a coherent beam of acoustic phonons. We propose a device based on undulated electrons in TBG, which we dub the phaser. The device generates phonon beams in a terahertz (THz) frequency range, which can then be used to produce THz electromagnetic radiation. The ability to generate coherent phonons in solids breaks new ground in controlling quantum memories, probing quantum states, realizing non-equilibrium phases of matter, and designing new types of THz optical devices.

\end{abstract}

\maketitle

\paragraph{Introduction.---}

Controlling and manipulating phonons is a long-sought goal offering a multitude of applications in electronics, information processing, and material science~\cite{Forst2011,Li2012,Balandin2012,Maldovan2013,Balandin2020,Subedi2014,Mankowsky2016,Juraschek2017a}, known as phononics. Recently, high-amplitude beams of phonons were employed to induce superconductivity~\cite{Mankowsky2014,Mitrano2016,Knap2016,Komnik2016,Babadi2017,Kennes2017,Cantaluppi2018,Liu2020} and to control ferroelectricity~\cite{Mankowsky2017,Nova2019,Li2019,Shin2020,Abalmasov2020} and magnetism~\cite{Nova2016,Juraschek2017,Shin2018,Radaelli2018,Maehrlein2018,Gu2018,Khalsa2018,Fechner2018,Juraschek2019,Disa2020,Juraschek2020a,Rodriguez-Vega2020,Juraschek2020,Juraschek2021,Stupakiewicz2021,Afanasiev2021}. Developing reliable sources of phonons is therefore of key importance for future advances in the field of phononics. Generation of coherent phonons in solids can be achieved through pumping by intense laser and magnetic fields~\cite{Miranda1976,Nunes1984,Nunes1998,Tronconi1986,Kittlaus2016}, or by the acoustic Cherenkov effect~\cite{Komirenko2000,Spector1962,Komirenko2003,Shinokita2016}.

A laser of phonons (i.e., a device for amplification of sound waves by stimulated emission) can serve as an efficient source of strong coherent acoustic waves with a narrow linewidth. Such devices were realized in the low-frequency range, from radio to gigahertz, in trapped ions~\cite{Wallentowitz1996,Vahala2009,Ip2018}, optical tweezers~\cite{Pettit2019}, nanomechanical resonators~\cite{Grudinin2010,Jing2014,Chafatinos2020,Lu2017,Wang2017,Zhang2018}, and magnetic systems~\cite{Bargatin2003,Ding2019}. A coherent amplification of terahertz (THz) phonons, yet below a threshold, was recently demonstrated in semiconductor superlattices~\cite{Beardsley2010}, and pump and probe experiments in silicon
carbide \cite{Cartella2018}.

\begin{figure}
  \centering
  \includegraphics[width=8.6cm]{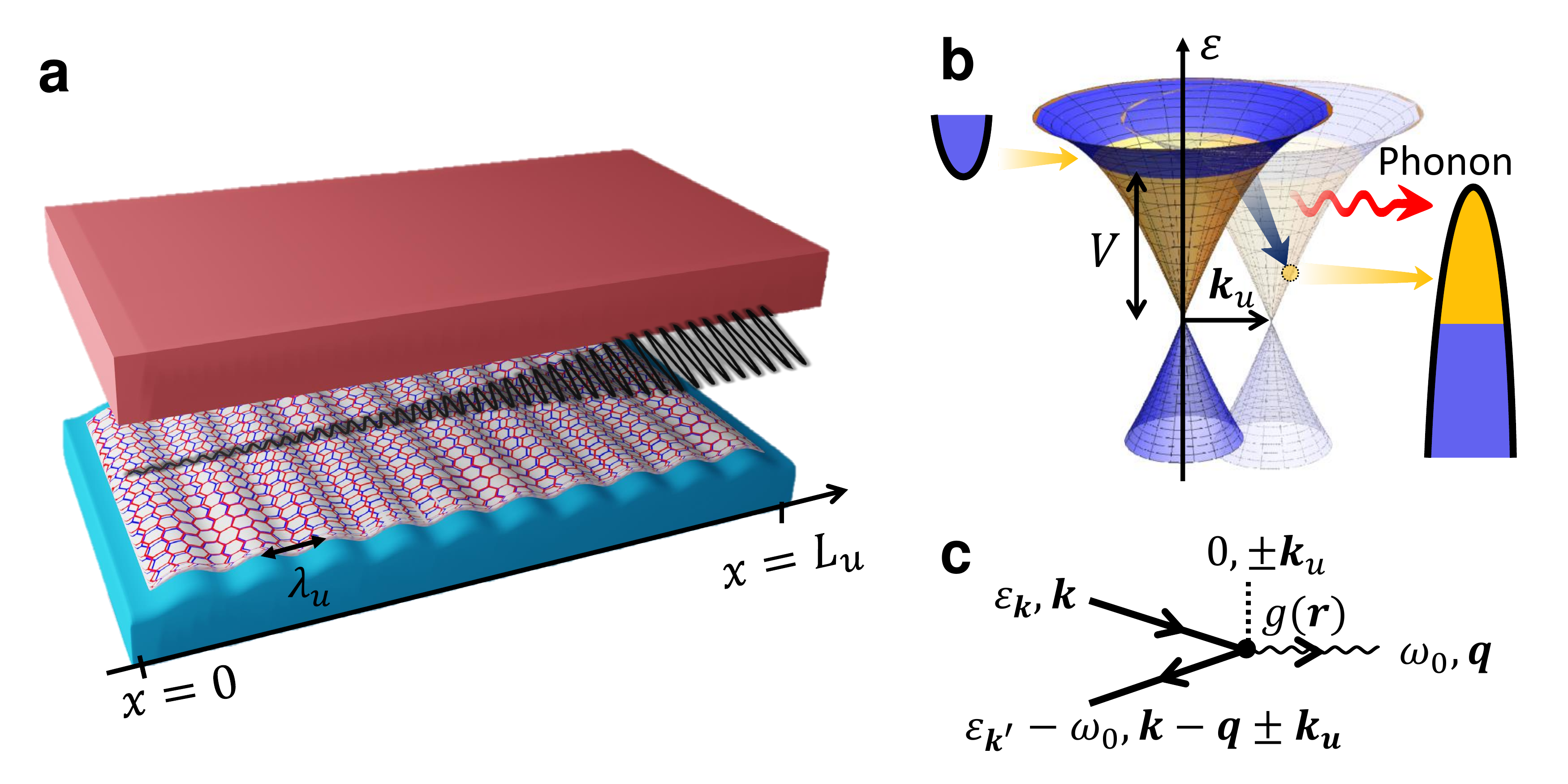}\\
  \caption{ \tb{The setup and the key  phonon-emission process.}
  \tb{a} Schematic illustration of the device. A layer of TBG is encapsulated between two gates of doped semiconductors. A nano-undulator is realized by a modulated in space uniaxial strain. The resonant frequency is controlled by the wavelength of the periodic structure, $\lm_u$.
  \tb{b} Illustration of the key phonon-emission process. The  blue areas indicate occupied states and the orange area  indicates the depleted region in the range $\ve=[0,V]$. The population inversion is imposed by the leads indicated by parabolas. 
  Phonon emission resonances occur between two replicas of the bands shifted by the wavevector $\vk_u=2\p/\lm_u\hat\vx$, induced by the nano-undulator. \tb{c}
Diagrammatic representation of a phonon emission process. The solid lines indicate the incoming and outgoing electrons, and the wiggly line indicates an emitted phonon. The dashed line indicates an extra crystal momentum provided by the modulation of the nano-undulator. In the slow-electron regime, the energy and crystal momentum can be only conserved in the presence of the momentum shift $\vk_u$. 
   \label{fig:Device}}
\end{figure}

Here, we present a model of a device for controlled amplification of acoustic phonons in the THz range, based on the newly-discovered narrow-band materials. We show that the unique bandstructures of such materials facilitate the amplification of coherent phonons in a narrow linewidth with low losses to incoherent modes. Furthermore, acoustic phonons have a long lifetime, giving rise to a high-gain and low-loss device~\cite{Ghosh2008}. 
Although phonon lasers are often referred to as ``sasers''~\cite{Zavtrak1997}, we dub our narrow-band-based device a ``phaser'', to highlight the quantum nature of the underlying mechanism.

Narrow-band materials were recently discovered in twisted bilayer graphene (TBG) ~\cite{Bistritzer2011,Cao2018,Cao2018a} and other moir\'e heterostructures~\cite{Ajayan2016,Balents2020}. In the TBG, the electronic dispersion can be tuned by a variation of the twist angle, reaching nearly flat bands at the ``magic'' twist angle. At the same time, the spectrum of the acoustic phonons of the TBG near the magic angle is almost unaffected by the twist angle~\cite{Koshino2019,Koshino2020,Ishizuka2021}, giving rise to the  ``\ti{slow-electron}'' regime in which the speed of sound surpasses the electronic group velocities~\cite{Sharma2021}. In this regime, the spontaneous emission of incoherent acoustic waves is suppressed by kinematic constrains.

We describe the realization of the device in a TBG tuned close to the magic angle and weakly modulated in space by a periodic uniaxial strain or a periodic array of screening gates, see Fig.~\ref{fig:Device}a. The periodicity of the modulation defines the resonant phonon mode of the phaser. Remarkably, for lasing in the THz range, the modulation wavelength should be in the mesoscopic scale. The electronic population inversion, necessary for the gain, is imposed by the external leads, in a structure similar to semiconductor laser diodes~\cite{Ammon1988}.

\paragraph{Toy model.---}
To develop an intuition for the lasing mechanism of the phaser, we begin by analyzing a toy model. Later, we numerically analyze the full band structure of the TBG, whose physics near the charge neutrality point can be described by this toy model. 
Yet, this model applies to more generic two-dimensional lattices in the slow-electron regime, with low-energy physics given by the Dirac Hamiltonian
\Eq{
\cH_\yD(\vk)=\hb v_\ye \vk\cdot \vsi,
\label{eq:ToyModel}
}
where $\vk=(k_x,k_y)$ is the in-plane crystal momentum, $\vsi=(\s^x,\s^y)$ is a vector of Pauli matrices acting in the pseudospin basis, and $v_\ye>0$ is the electronic group velocity. Eq.~\eqref{eq:ToyModel} is diagonalized by the Bloch states $\frac{e^{i\vr\cdot\vk}}{\sqrt{\cA}}\ket{\y_{\vk\a}(\vr)}$, where $\ket{\y_{\vk\a}(\vr)}$ is periodic in the unit cell, corresponding to the eigenvalues
$\ve_\a(\vk)=\a\hb v_e|\vk|$, where $\a=\pm$ and $\cA$ is the area of the system. The eigenstates are created by the  operators $\hat c\dg_{\vk\a}$. In the toy model, we assume no spin or pseudospin degrees other than $\vsi$ (additional degrees of freedom such as valley, spin, and layer indices of the TBG would not qualitatively change the effect).

We consider a regime in which $v_\ye<c_{\rm ph}$, where $c_{\rm ph}$ is the speed of sound in the material, assumed to be uniform and isotropic. 
The corresponding sound waves are described by the lattice displacement operator $\hat \vu(\vr,t)=(\hat u_x,\hat u_y)$\footnote{We do not consider the flexural modes of suspended graphene.}, which can be expanded in the eigenmodes $\hat \vu(\vr,t)=\frac{1}{\sqrt{\cA}}\sum_\vq e^{i\vq \cdot\vr-i\w_l(\vq)t} \vc_{l}(\vq)\tilde u_l(\vq)$. Here, $\vr=(x,y)$, and $\vc_l(\vq)$ is the unit vector denoting the direction of the displacement in the mode $l$ and crystal momentum $\vq$. 
Focusing on the lowest energy acoustic mode with $l=0$, we assume a dispersion $\w_0(\vq)=c_{\rm ph}|\vq|$, and coupling to electrons 
\Eq{
\hat \cH_{\rm ep}=\int d^2\vr g(\vr) \hat \cO_{ij}(\vr) \dpa_i \hat u_j(\vr).
\label{eq:ToyModel_ElPh}
}
Here, $\hat\cO_{ij}(\vr)$ is a local electronic operator with $i,j=\{x,y\}$ and $g(\vr)$ denotes the coupling strength assumed to be non-uniform in space. The spatial dependence of $g(\vr)$ is specified below.

The system is connected to two external leads imposing population inversion for the electrons. The two leads are electron- and hole-doped semiconductors, with the bottom edge of the conduction band of the electron-doped semiconductor and the top edge of the valence band of the hole-doped semiconductor set at the energy $\ve=V$. The chemical potential of the electron-doped semiconductor is set at $\ve=\ve_{\rm top}$, corresponding to  the top of the upper band of the TBG [denoted by $\a=+$, see below Eq.~\eqref{eq:ToyModel}]. 
The chemical potential of the hole-doped semiconductor is set to the charge neutrality point of the TBG, $\ve=0$. 
For simplicity, we assume that 
the tunneling rate of the electrons between the system and the leads is faster than the decay rate of the electrons in the system due to relaxation and phonon emission processes~\footnote{Otherwise one needs to compute the full electronic steady state in the presence of the tunneling from the leads and relaxation, which we leave for future investigation}. With this assumption and for \ti{zero-temperature} leads, the occupation probability $f_{\a\vk}=\av{\hat c\dg_{\a\vk} \hat c_{\a\vk}}$ of the electrons can be approximated by $f_{\a\vk}=0$ for $0<\ve_\a(\vk)<V$, and $f_{\a\vk}=1$ otherwise, imposing an inverted population in the bottom of the upper band, see Fig.~\ref{fig:Device}b.

Such an inverted population is virtually decoupled from the phonons to the leading order in the electron-phonon coupling when this coupling is \ti{spatially uniform}. 
This is because, in the ``slow-electron'' regime, it is impossible to simultaneously conserve energy and crystal momentum in a single-phonon emission. 
Therefore, in this case, the incoherent phonon background field created by the non-equilibrium electronic state is suppressed compared to wide-band materials. 
This virtual decoupling between the electrons and the phonons provides an important baseline condition for lasing. However, the electrons should be coupled to at least one phonon mode, to generate a coherent beam.

Following the concept of free-electron lasers~\cite{Madey2003,Pellegrini2016,Friedman1988,Roberson1998}, emission in a selected mode can be induced by spatially modulated electron-phonon coupling coefficient, $g(\vr)$. 
In what follows, we consider a coupling modulated along the $\hat \vx$ direction, with a wavelength $\lm_u=2\p/k_u$, and uniform along $\hat \vy$, $g(\vr)=g_0+2g_1 \cos(k_u x)$, see Fig.~\ref{fig:Device}a for illustration. We denote the region of the system where $g_1\ne 0$, a \ti{nano-undulator}, by analogy with a magnetic undulator in free-electron lasers. 
The physical realization of the nano-undulator in the TBG is discussed below.

In the nano-undulator,  the conservation of crystal momentum in a phonon emission process obeys $\vk'-\vk=\vq+n  k_u\hat\vx$, where $\vk$ and $\vk'$ are respectively the crystal momenta of the electron before and after the emission of a phonon with momentum $\vq$, and $n=\{-1,0,1\}$ [see Fig.~\ref{fig:Device}c]. The additional momentum shift arises from the expansion of $g(\vr)$  in its spatial Fourier components, $g(\vr)=\sum_{n} g_n e^{i n k_u x}$, where $g_1=g_{-1}$. The phase shift of the electron-phonon components corresponds to a momentum shift in Eq.~\eqref{eq:ToyModel_ElPh}. In turn, the energy conservation is not affected by the static modulation of the coupling and reads $\ve_{+}(\vk')-\ve_+(\vk)=\hb\w_0(\vq)$.  
For phonons propagating in the $\hat \vx$-direction, the energy and crystal momentum conservation is satisfied by two resonant modes, with frequencies
\Eq{
\w_{R\pm}=\frac{c_{\rm ph}k_u}{(c_{\rm ph}/v_\ye)\pm 1}.
\label{eq:ResonantFrequency}
}
This formula demonstrates  that the resonant frequency of the laser can be controlled by tuning the modulation wavevector of the nano-undulator, $k_u$.

\paragraph{Estimation of the gain.---}

To estimate the gain of the device, we consider a coherent sound wave incident at $x=0$ with amplitude $\vu_0$ and frequency $\w$, propagating in the positive $\hat\vx$ direction, see Fig.~\ref{fig:Device}a. This sound wave can be generated i.e., by seeding from an external source or by spontaneous emission processes. 
We parametrize the amplification of the sound wave in the nano-undulator by an exponential factor with the gain coefficient $\g_\w$~\cite{Ammon1988} \Eq{
\vu_\w(\vr,t) =  \vu_0  e^{\g_\w x}e^{i(qx-\w t)},
\label{eq:AcousticMode}
}
where $q=\w/c_{\rm ph}$.
Such a sound wave, after propagating through the nano-undulator, carries a period-averaged power density of
\Eq{
\cP_\ys(\w)=\frac {1} {2a^2 L_u} c_{\rm ph}M \w^2|\vu_0|^2(e^{2\g_\w L_u}-1),
\label{eq:PhononGain}
}
where $L_u$ is the nano-undulator length along the $\hat \vx$ direction, $M$ is the mass of the atoms comprising the lattice, and $a$ is the lattice constant.

\begin{figure}
  \centering
  \includegraphics[width=8.6cm]{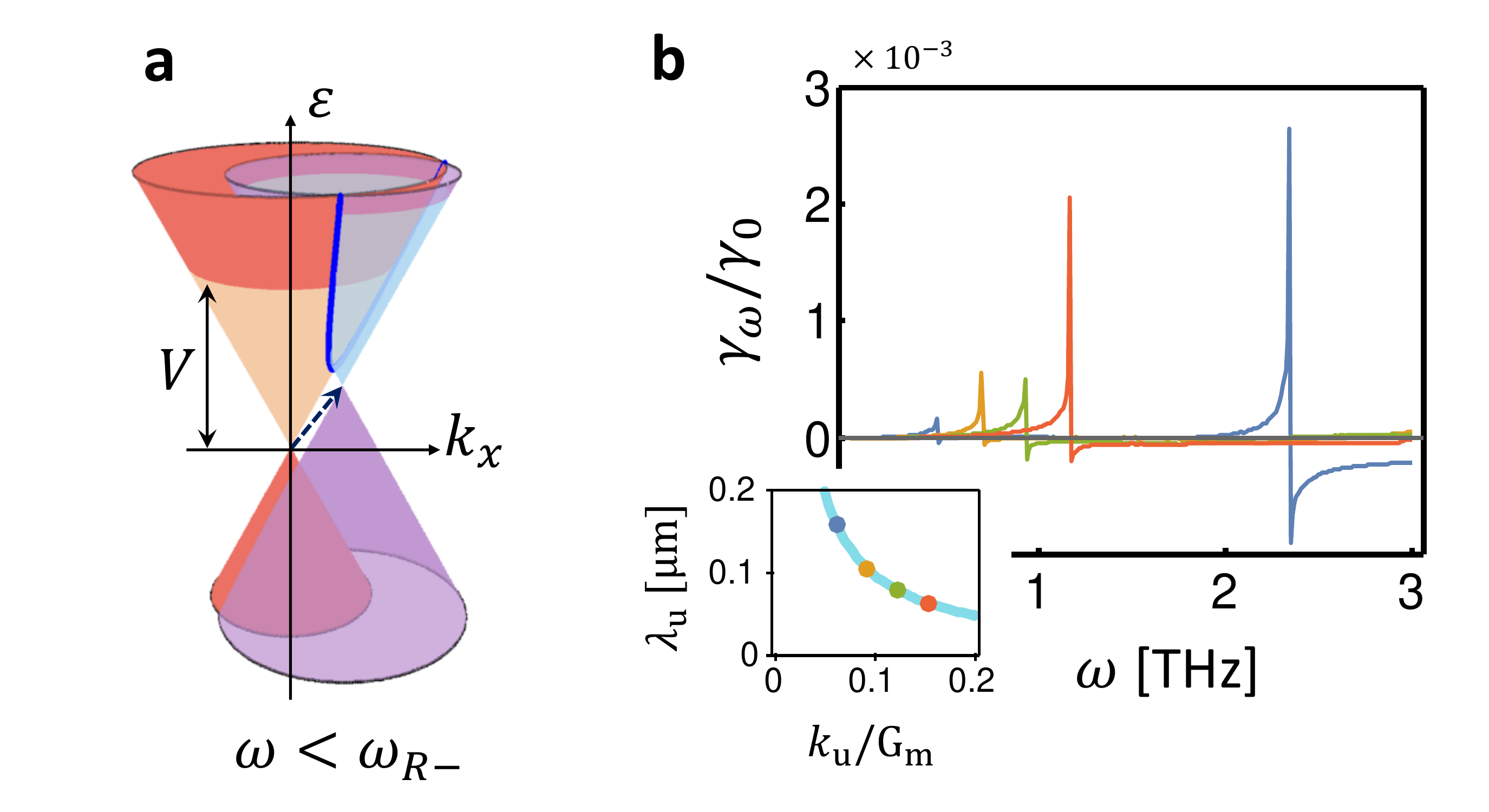}\\
  \caption{ \tb{The gain in the toy model.} 
  \tb{a} Dirac dispersion of the electrons described by the toy model [see Eq.~\eqref{eq:ToyModel}] and the one shifted by the energy $\w$ and crystal momentum $q+k_u$ in the $\hat\vx$ direction (indicated by the dashed arrow). The phonon emission rate is proportional to the intersection area of the two Dirac dispersions [see Eq.~\eqref{eq:Power_ep}], indicated by the blue line. \tb{b} The 
gain of the device normalized by $\g_0$ (see Eq.~\eqref{eq:Gain}) as a function of the phonon frequency calculated in the toy model, for four values of $k_u$. The values of $k_u$ and the corresponding wavelengths $\lm_u=2\p/k_u$ are indicated in the inset. The gain exhibits resonance peaks at the frequencies given by Eq.~\eqref{eq:ResonantFrequency}.    \label{fig:GainAn}}
\end{figure}

This power is the result of  the system steadily emitting acoustic phonons in the nano-undulator. 
In the low-gain limit, the period-averaged power density emitted by the electrons stimulated by the field $\vu_\w$, can be found using Fermi's golden rule 
\footnote{See supplemental material for derivation.},
\EqS{
\cP_{\ye}(\w) =\frac{2\p\w}{\cA}\sum_{\a\a',\vk\vk'}&f_{\a\vk}[\cM^{\a\a'}_{\vk\vk'}\dl(\ve_{\a}(\vk)-\ve_{\a'}(\vk')-\hb\w)-\\
&-\cM^{\a\a'}_{\vk'\vk}\dl(\ve_{\a}(\vk)-\ve_{\a'}(\vk')+\hb\w)].
\label{eq:Power_ep}
}
Here, ($\a,\vk$) and ($\a',\vk'$) respectively denote the electronic band and crystal momentum before and after the interaction with the acoustic wave and we defined $\cM^{\a\a'}_{\vk\vk'}=|\cA\inv \int d^2\vr e^{i\vr(\vk-\vk')}\braoket{\y_{\a\vk}(\vr)}{\hat \cH_{\rm ep}}{\y_{\a'\vk'}(\vr)}|^2$, where the integration is over the area of the nano-undulator.
We note that coherent phonon generation in a more generic case can be studied by analyzing electron-phonon instabilities of the equations of motion, as is outlined in the supplementary material.
Given, that the sound wave is coherent, we can approximate the acoustic field operator in the expression for $\hat \cH_{\rm ep}$ [Eq.~\eqref{eq:ToyModel_ElPh}] by its expectation value $\av{\hat \vu}\eqa \vu_\w$, given in Eq.~\eqref{eq:AcousticMode}. Then we obtain $\cM^{\a\a'}_{\vk\vk'}=q^2|\vu_0|^2\sum_n |\frac{g_n}{\cA}|^2   |\int d^2\vr e^{\g_\w x}  e^{i\vr(\vk-\vk'-(q+n k_u)\hat\vx)}    \av{\hat\cO}^{\a\vk}_{\a'\vk'}  |^2$, where $\av{\hat\cO}^{\a\vk}_{\a'\vk'}=\int d^2\vr \braoket{\y_{\a\vk}}{\hat \cO_{x i}(\vr)c_i}{\y_{\a'\vk'}}$, and $c_i$ is the $i$-th component of the unit vector pointing in the direction of $\vu_0$.

Assuming $\av{\cO}$ does not have a strong momentum dependence, and taking a small gain limit $\g_\w\to0$,  $\cM^{\a\a'}_{\vk\vk'}$ is non-zero only when $|\vk-\vk'-(q+nk_u)\hat\vx|^2<(2\p)^2/\cA$. 
In the thermodynamic limit ($\cA\to \infty$), the values of $\vk$ contributing to the sum in the expression for $\cP_\ye(\w)$ in  Eq.~\eqref{eq:Power_ep}, lie near the intersection line of two cones described by $\ve_\a(\vk)$ and $\ve_{\a'}(\vk')+\hb\w$, where $\vk'=\vk+(q+nk_u)\hat\vx$, and $f_{\a \vk}\ne f_{\a' \vk'}$, see Fig.~\ref{fig:GainAn}a. The largest value of $\cP_\ye(\w)$ is obtained for $\w\eqa \w_{R,n}$, where $n=\pm$ [see Eq.~\eqref{eq:ResonantFrequency}], where the two cones are tangent. The area in the momentum space where the two cones are nearly tangential diverges as $\dl \w_n^{-\half}$, with $\dl\w_n=\w_{R,n}-\w$, as $\w$ approaches $\w_{R,n}$ from below, giving rise to a resonance peak in $\cP_\ye(\w)$. For $\w>\w_{R,n}$ the intersection line of the cone with $\a=+$ and the cone with $\a'=-$, contributes to a negative peak corresponding to the absorption of phonons. 

The gain in the system, $\g_\w$, can be found by setting $\cP_\ys(\w)=\cP_\ye(\w)$ and using the expressions of $\cP_\ys(\w)$ and $\cP_\ye(\w)$ as a function of $\g_\w$ [Eqs.~\eqref{eq:PhononGain} and \eqref{eq:Power_ep}]. In the small gain limit ($\g_\w\to 0$), we obtain
\Eq{
\g_\w=\cP^0_\ye(\w)a^2/(c_{\rm ph}M\w^2 |\vu_0|^2),
\label{eq:GainDefinition}
}
where $\cP^0_\ye(\w)=\cP_\ye(\w)|_{\g_\w=0}$. Estimating Eq.~\eqref{eq:Power_ep} in the limit $0<\dl \w_n\ll \w_{R,n}$ and $V\gg\hb\w_{R,n}$ \cite{Note3}, we find 
\Eq{
\g_\w=\g_0\sum_{n=\pm} \hb\w_{R,n}\sqrt{2\w_{R,n}/\dl \w_n}\cN_D(V)a^2,
\label{eq:Gain}
}
where $\g_0=  g_1^2\av{\cO}^2 /(\hb c_{\rm ph}^3M)$ and $\cN_\yD(\ve)=\ve/(2\p\hb^2v_\ye^2)$ is the density of states of the Dirac dispersion. 
Fig.~\ref{fig:GainAn}b shows $\g_\w$ as a function of $\w$ for a few values of $k_u$ shown in the inset of this figure.

\paragraph{Realization of the phaser in the TBG.--- \label{sec:TBGRealization}}
Having established the lasing mechanism for the toy model, we now discuss its realization in the TBG. The TBG consists of two graphene monolayers twisted by a relative angle $\q$, giving rise to a moir\'e supper lattice with the lattice constant \cite{Andrei2020} $a_\ym=a/(2\sin(\q/2))$. For small twist angles, the dispersion near the charge neutrality exhibits narrow bands, whose low-energy physics can be approximated by Eq.~\eqref{eq:ToyModel}  with additional degenerate spin and valley quantum numbers~\cite{Bistritzer2011,Koshino2018}. For concreteness, we focus on $\q=1.4^\circ$, where we found $v_\ye\eqa 2\times 10^6  \, {\rm cm/sec}$, which is below the speed of sound in the material, approximated here by  $c_{\rm ph}=3\times 10^6 \, {\rm cm/sec}$.
We consider two alternative realizations of a nano-undulator, through spatial mesoscopic modulation of the electron-phonon coupling [see Eq.~\eqref{eq:ToyModel_ElPh}]~\footnote{A spatially modulated perturbation also modifies the electron and phonon dispersions, by opening minigaps at $\pm \vk_u/2$, which do not affect phase space region relevant for lasing.}.

The first realization is based on a spatially modulated uniaxial strain, as illustrated in Fig.~\ref{fig:Device}a. Such a modulation can be realized by placing the TBG on an architected nanostructure or by applying temperature gradients~\cite{Deng2016,Frisenda2017,Ludacka2018,Banerjee2020,Hsu2020}. 
Weak spatially periodic strain modulates the lattice geometry of each graphene monolayer, which in turn modulates the electron-phonon coupling~\cite{Bi2019,Koshino2020}. For a strain applied along the $x$-direction of the form $\e_0\cos(k_u x)$, the spatially-modulated part of the electron-phonon coupling in each monolayer can be expressed by Eq.~\eqref{eq:ToyModel_ElPh} with  $g_1=\frac{\sqrt{3}}{4a}\hb v_F \be\e_{0}$, and $\hat \cO_{ij}(\vr)=(\hat c\dg_{\vr,A}\hat c_{\vr,B}+\hat c\dg_{\vr,B}\hat c_{\vr,A})(\dl_{i,x}\dl_{j,x}-\dl_{i,y}\dl_{j,y})$, where $\hat c\dg_{\vr,A/B}$ creates an electron in the sublattice A or B of the graphene monolayer at the unit cell located at $\vr$ \cite{Note3}. 
For $\e_0\eqa 5\%$ strain, we estimate $g_1\eqa 0.15\, {\rm eV}$.

The second realization is based on a periodic array of metallic gates at distance $d$ from the TBG. The gates change the screening efficiency of the Coulomb interaction between the electronic charge density and the lattice ions \cite{Mariani2010,Hwang2007,VonOppen2009,Suzuura2002a,Manes2007,Sohier2014,Park2014}. For a phonon of momentum $q$, we approximate the renormalized coupling term by $g(\vr)=D_0 q/[q+q_{\rm TF}\tanh(q d(\vr))]$, where $d(\vr)$ is the distance from the gates, periodically changing between $d(\vr)\eqa d$ when $\vr$ above one of the gates and $d(\vr)\to\infty$ when $\vr$ is in the space between the gates; $q_{\rm TF}$ is the Thomas Fermi wavevector and $D_0$ the bare electron-phonon coupling.
To approximate the electron-phonon coupling by Eq.~\eqref{eq:ToyModel_ElPh}, in the limit $q_{\rm TF}\gg q, 1/d$, we estimate $g_1\eqa \frac{1}{4}\frac{D_0}{1+q_{\rm TF }d}$ and $\hat\cO_{ij}(\vr)\sim \hat\ro(\vr)\dl_{ij}$, where $\hat\ro(\vr)=\hat c\dg_{\vr,A}\hat c_{\vr,A}+\hat c\dg_{\vr,B}\hat c_{\vr,B}$ measures the density. For $q_{\rm TF} d\eqa 3$~\cite{Kim2020} and $D_0= 50\, {\rm eV}$, we estimate $g_1\eqa 3\, {\rm eV}$. For this value of $g_1$, we obtain $\g_0\eqa 8.5\, {\rm \m m }\inv$.

\begin{figure}
  \centering
  \includegraphics[width=8.6cm]{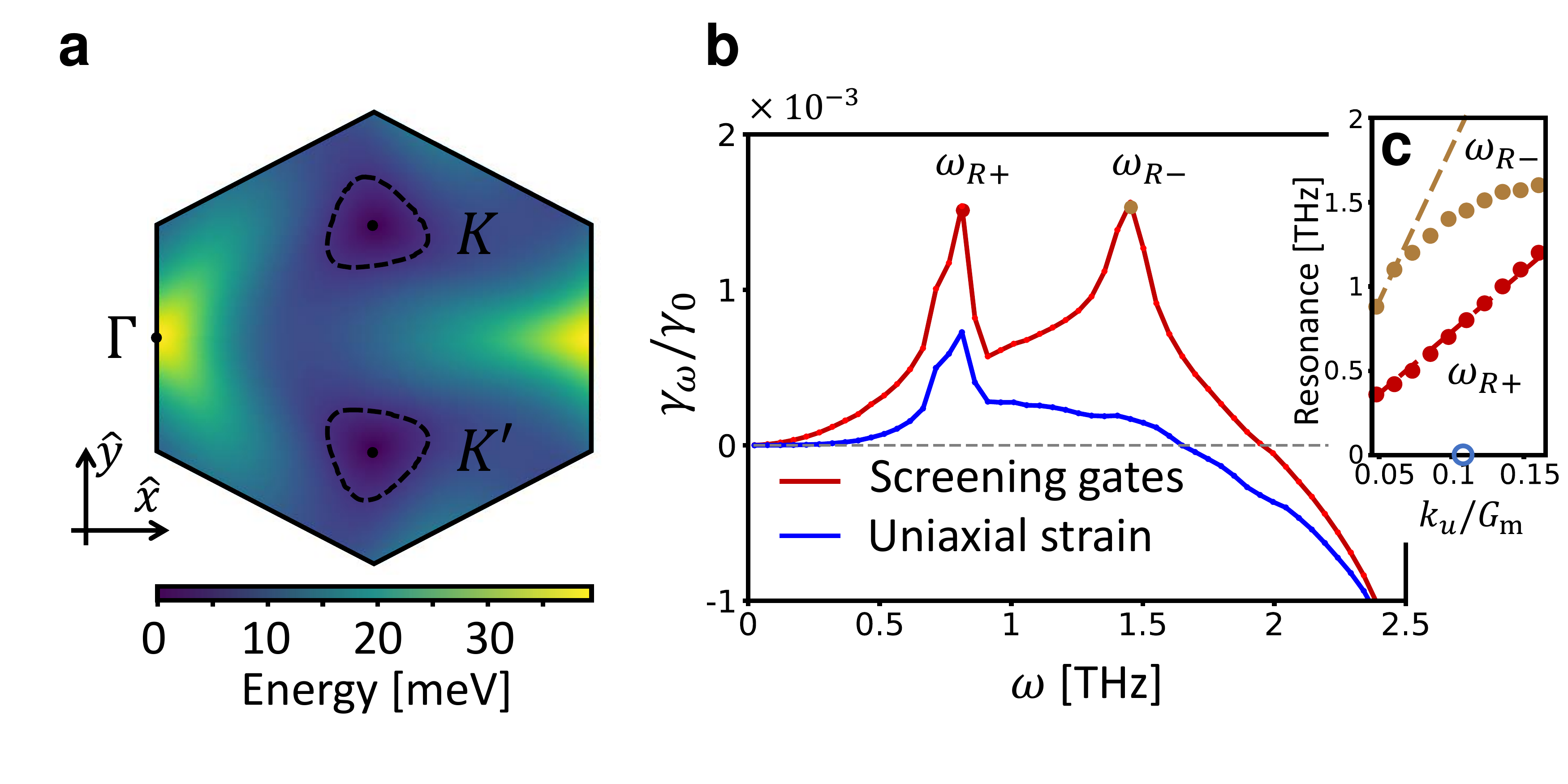}\\
  \caption{ \tb{Numerical analysis of the phaser based on the TBG.} \tb{a} The band structure of the upper band of the TBG near the charge neutrality point for a single valley, shown on a mini Brillouin zone centered around the Dirac points, $K$, and $K'$. Curves around $K$ and $K'$ points indicate  the equipotential lines for $\ve=6\, {\rm meV}$. \tb{b} The gain calculated by Eqs.~\eqref{eq:Power_ep} and \eqref{eq:GainDefinition} for the wavefunctions of the TBG  for the realization of the nano-undulator by a modulated uniaxial strain and an array of screening gates, for the wavevector $k_u\eqa 0.11 G_{\ym}$ indicated by a circle on the x-axis in panel c. \tb{c} The frequency of the resonant peaks of the gain as a function of $k_u$. Dashed lines indicate the resonances predicted by the toy model [Eq.~\eqref{eq:ResonantFrequency}]. \label{fig:Numerical}}
\end{figure}

\paragraph{Numerical analysis of the gain.---}
We numerically simulated the band structure of the TBG using the continuum model~\cite{Bistritzer2011, Koshino2018}. Our goal was to verify that the full bandstructure of the TBG exhibits resonance peaks for phonon emission, as predicted by the toy model, and to compare their frequencies to  Eq.~\eqref{eq:ResonantFrequency}. Fig.~\ref{fig:Numerical}a shows the spectrum of the upper band of the TBG in a single valley in the mini-Brillouin zone centered around the Dirac points, see details in the supplementary material. The contours surrounding the $K$ and $K'$ points indicate the equipotential curves $\ve=V$.

We evaluated the gain in the full model of the TBG using Eq.~\eqref{eq:GainDefinition}, directly estimating $\cP_\ye(\w)$ by  Eq.~\eqref{eq:Power_ep} in the two realizations of the nano-undulator, described above.
Fig.~\ref{fig:Numerical}b shows the numerically obtained gain as a function of the frequency for $k_u\eqa 0.11G_\ym$, where $G_\ym=2\p/a_\ym$, for the two realizations of the nano-undulator. Both curves exhibit a peak near $\w\eqa 0.8\, \rm THz$, corresponding to the analytical estimate of $\w_{R+}$ [defined in Eq.~\eqref{eq:ResonantFrequency}]. The curve corresponding to an array of screening gates exhibits an additional peak near $\w\eqa 1.5\, \rm THz$, which is missing in the curve of the uniaxial strain, due to the selection rules, suppressing the transitions for large phonon frequencies.

Fig.~\ref{fig:Numerical}c shows the frequencies of the two resonances of the gain for the screening-gates realization of the nano-undulator as a function of $k_u$. Dashed lines denote the analytical prediction of $\w_{R\pm}$. The numerical curve of the low-frequency peak approximately coincides with the analytical curve corresponding to $\w_{R+}$ as a function of $k_u$ with $v_\ye=2\times 10^6 \, \rm cm/sec$ [see Eq.~\eqref{eq:ResonantFrequency}]. The position of the second peak does not show a linear dependence as a function of $k_u$, as predicted by Eq.~\eqref{eq:ResonantFrequency}, due to deviations of the TBG band structure from the linear dispersion [Eq.~\eqref{eq:ToyModel}] for high phonon energies. We can fit its resonance frequency at $k_u/G_\ym= 0.05$, by $\w_{R-}$ with $v_\ye=1.5\times 10^6 \, \rm cm/sec$.

\paragraph{Lasing threshold.---}

To lase, the device should reach the lasing threshold, i.e., the gain should exceed the loss. The loss of phonons mostly occurs due to electron-phonon, phonon-phonon, and impurity scattering. The lifetime of acoustic phonons in clean graphene can reach $\ta_{\rm ph}\eqa0.3\, \rm \m sec$, for long-wavelength phonons~\cite{Bonini2012}, which results in $\g_{\rm loss}=(\ta_{\rm ph}c_{\rm ph})\inv\eqa 2\times 10^{-4}\, \rm \m m\inv$. This value is below the gain of the system, estimated slightly above the resonance peak.

To have a sufficient gain, the system can be placed in an acoustic cavity. Such cavities were realized, e.g.,  in Ref.~\onlinecite{Maryam2013}. The phonon-loss in a cavity is given by $\g_{\rm cavity}=-\log(R_1 R_2)/(2L_u)$, where $R_1$ and $R_2$ are the reflectivities of the two mirrors. For $R_1R_2=0.97$ and $L_u\eqa 5\, \rm \m m$, we obtain $\g_{\rm cavity}\eqa 0.001\, \rm \m m\inv$. This results in a Q-factor of the cavity for the phonons of about $Q\sim 10^5$.

\paragraph{Discussion.---}
In this manuscript, we presented a model of a phonon laser device based on the ``narrow-band'' regime, dubbed a phaser. 
The phaser generates coherent phonon beams in the THz range. We demonstrated two realizations of the phaser in the TBG tuned near the magic angle, with a spatially modulated uniaxial strain and an array of screening gates [see Fig.~\ref{fig:Device}a]. The periodicity of the structure can control the resonant frequency of the device.
The phaser opens up new avenues in driving the TBG into a non-equilibrium regime through moir\'e Floquet engineering~\cite{Katz2020,Rodriguez-Vega2021}, extending the driving sources to THz frequencies and finite momenta~\cite{Hubener2018}. 

The lattice oscillations caused by the phonon beam are coupled to plasmon modes through Coulomb interactions and the electron-phonon coupling. 
The resulting charged modes generate a THz electromagnetic field evanescent in the direction perpendicular to the TBG plane. 
We estimate the electric field amplitude near the surface \cite{Note3} by 
$|\vec E|=2\sqrt{2}\p e\ro_0 \lm q^2 |\av{\hat\vu}|$. Here, $\lm q$ denotes the relative charge fluctuation which we estimate as $\lm q \eqa 2\times 10^{-2}$, and $\ro_0$ is the electronic density taken as $\ro_0\eqa 1/a_\ym^2$. Assuming that the phaser in the saturation regime creates lattice waves of the order of $|\av{\hat\vu}|\eqa 0.1 a$, we estimate, $|\vec E|\eqa 30 \, \rm kV/m$.
Such an electric field can be detected by placing a dipole antenna near the surface of the TBG. An oscillating evanescent electric field can be transformed into THz electromagnetic radiation, through a meta-material structure. We leave the  analysis of this problem for future studies.

In our analysis, we focused on the single-particle electronic bands of the TBG. In the presence of the electron-electron interactions, the Fermi velocity may be renormalized, yet the slow-electron regime can be still achieved~\cite{Goodwin2020}. Furthermore, the Dirac dispersion near the charge neutrality point is protected by the $C_2\cT$ symmetry (two-fold rotation times time-reversal) and will be preserved unless it is spontaneously broken~\cite{Po2018}.

The toy model of the phaser [Eq.~\eqref{eq:ToyModel}] can be realized in other experimental platforms.  For example, a ``slow-band'' regime can be realized in cold atoms, using Bose-Fermi mixtures \cite{Illuminati2004,Wang2005}. We note, however, that the energy scales of cold atom setups are a few orders of magnitude smaller than in solids, giving rise to a different range of resonant frequencies.

\begin{acknowledgments}
We thank Kenneth Burch, Jerome Faist, Mohammad Hafezi, Atac Imamoglu, Cyprian Lewandowski, Marios Michael, Leo Radzihovsky, and Christopher Yang for valuable discussions. 
G. Refael and I. Esin are grateful for support from the Simons Foundation and the Institute of Quantum Information and Matter, as well as support from the NSF DMR grant number 1839271. 
E. Demler and I. Esterlis acknowledge support from the ARO grant ``Control of Many-Body States Using Strong Coherent Light-Matter Coupling in Terahertz Cavities''.
This work is supported by ARO MURI Grant No. W911NF-16-1-0361, and was performed in part at Aspen Center for Physics, which is supported by National Science Foundation grant PHY-1607611.
\end{acknowledgments}

\bibliography{Bibliography.bib}

\end{document}


\title{Generating coherent phonon waves in narrow-band materials: a twisted bilayer graphene phaser - Supplementary Material}

\maketitle

\section{Phonon instability from the equation of motion formalism}

In this section, we derive the equations of motion for the electronic and bosonic operators and demonstrate an instability of the resonant phononic mode. For simplicity, here and throughout these notes we work in the units in which $\hb=1$.
We consider the electronic Hamiltonian 
\Eq{
\hat\cH_\yD=\sum_{\vk,\a=\pm} \ve_\a(\vk)\hat c\dg_{\vk\a}\hat c_{\vk\a},
}
where $\ve_\a(\vk)=\a v_\ye|\vk|$.
Similarly the Hamiltonian for the acoustic phonons reads
\Eq{
\hat\cH_{\rm ph}=\sum_\vq \w_0(\vq) \hat b\dg_\vq \hat b_\vq,
}
where $\w_0(\vq)=c_{\rm ph}|\vq|$. The electron-phonon coupling is given by 
\Eq{
\hat\cH_{\rm ep}=\sum_{\vk,\vq,n} g_{n}[\hat c\dg_{\vk,\a}\hat c_{\vk+\vQ_n,\a'}\hat b_\vq+\hat c\dg_{\vk,\a}\hat c_{\vk-\vQ_n,\a'}\hat b_\vq\dg],
}
where $\vQ_n=\vq+n\vk_u$ and $n=\{-1,0,1\}$.
The full Hamiltonian of the system is given by the sum, $\hat\cH=\hat \cH_{\yD}+\hat\cH_{\rm ph}+\hat\cH_{\rm ep}$. 
The equations of motion for the operators in the Heisenberg picture are given by $-i \dpa_t \hat c_{\vk\a}=[\hat\cH,\hat c_{\vk\a}]$ and $-i \dpa_t \hat b_{\vq}=[\hat\cH,\hat b_{\vq}]$. Explicit calculation of the commutator yields
\EqS{
-i\dpa_t\hat c_{\vk\a}=-\ve_{\a}(\vk)\hat c_{\vk\a}-
\sum_{\vq,n} g_{n}[\hat c_{\vk+\vQ_n,\a'}\hat b_\vq+\hat c_{\vk-\vQ_n,\a'}\hat b_\vq\dg].
\label{eq:EOMfermions}
}
The equation for $\hat c_{\vk\a}\dg$ is obtained from the complex conjugate. Next, we compute $-i\dpa_t\av{\hat c_{\vk\a}\dg\hat c_{\vk',\a'}}=-i\av{[\dpa_t\hat c_{\vk\a}\dg]\hat c_{\vk',\a'}}-i\av{\hat c_{\vk\a}\dg[\dpa_t\hat c_{\vk',\a'}]}$, which reads
\EqS{
&-i\dpa_t\av{\hat c_{\vk\a}\dg\hat c_{\vk'\a'}}=[\ve_\a(\vk)-\ve_{\a'}(\vk')]\av{\hat c_{\vk\a}\dg\hat c_{\vk'\a'}}+\\
&+g\sum_{\vq,n\be}[\av{\hat c_{\vk+\vQ_n\be}\dg\hat c_{\vk'\a'}}\av{\hat b_\vq}+\av{\hat c_{\vk-\vQ_n\be}\dg\hat c_{\vk'\a'}}\av{\hat b_\vq\dg}]-\\
&-g\sum_{\vq,n\be}[\av{\hat c_{\vk\a}\dg\hat c_{\vk'+\vQ_n\be}}\av{\hat b_\vq\dg}+\av{\hat c_{\vk\a}\dg\hat c_{\vk'-\vQ_n\be}}\av{\hat b_\vq}].
\label{eq:EOMOccupations}
}
Similarly,
\begin{subequations}
\begin{eqnarray}
&&-i\dpa_t \av{\hat b_\vq}=-\w_0(\vq)\av{\hat b_\vq}-\sum_{n,\vk} g_{n}\av{\hat c\dg_{\vk\a}\hat c_{\vk+\vQ_n,\a'}}\label{seq:EOMphonon1}\\
&&-i\dpa_t \av{\hat b_\vq\dg}=\w_0(\vq)\av{\hat b_\vq}+\sum_{n,\vk} g_{n}\av{\hat c\dg_{\vk+\vQ_n\a}\hat c_{\vk,\a'}}.
\end{eqnarray}
\label{eq:EOMphonon}
\end{subequations}
Next, we perform the transformation $-i\dpa_t\to\w$ and solve for the eigenmodes. To simplify this problem, we assume a single coherent phonon mode with momentum $\vq$ corresponding to $\vQ_n$, and set $\vk'=\vk+\vQ_n$ in Eq.~\eqref{eq:EOMOccupations}, yielding
\EqS{
\w\av{\hat c_{\vk\a}\dg\hat c_{\vk+\vQ_n,\a'}}=&[\ve_\a(\vk)-\ve_{\a'}(\vk+\vQ_n)]\av{\hat c_{\vk\a}\dg\hat c_{\vk+\vQ_n\a'}}+\\
&+g_n[f_{\vk+\vQ_n,\a'}-f_{\vk,\a}]\av{\hat b_\vq}.
\label{eq:EOMOccupations2}
}
Here, we defined $f_{\vk\a}=\av{\hat c_{\vk\a}\dg\hat c_{\vk\a}}$ and neglected expectation values with momentum $2\vQ_n$. Combining Eq.~\eqref{eq:EOMOccupations2} and  Eq.~\eqref{seq:EOMphonon1}, we obtain a linear equation. Imaginary eigenvalues correspond to an exponentially growing mode of phonons hybridized with electron-hole excitations.
An approximate solution can be obtained if we assume $[\ve_{\a'}(\vk+\vQ_n)-\ve_\a(\vk)]=\D \ve$, and $\a=\a'=+$. Then one can sum both sides of Eq.~\eqref{eq:EOMOccupations2} over $\vk$, leading to and eigenvalue equation
\EqS{
(\w+\D \ve)(\w+\w_0(\vq))=-g_n^2\sum_\vk\D f_\vk,
}
where $\D f_\vk=f_{\vk+\vQ_n,+}-f_{\vk,+}$ 
The solution reads
\Eq{
\w_{\pm}=\frac{\D\ve+\w_0(\vq)}{2}\pm\half\sqrt{(\D\ve-\w_0(\vq))^2-4g_n^2\sum_\vk\D f_\vk}.
}
Therefore, near the resonance the eigenvalues obtain an imaginary component, when $(\D\ve-\w_0(\vq))^2<4g_n^2\sum_\vk\D f_\vk$.

\section{Boltzmann equation from the equation of motion formalism}
Here, we derive the Boltzmann equation [Eq.~(6) in the main text] from the equations of motion, Eq.~\eqref{eq:EOMfermions} and Eq.~\eqref{eq:EOMphonon}. First, we transform the operators to the ``interaction picture'', $\hat c_{\vk\a}=e^{-i \ve_{\a}(\vk)t}\hat c^I_{\vk\a}$ and $\hat b_{\vq}=e^{-i \w_0(\vq)t}\hat b^I_{\vq}$. Next, we integrate both sides of Eq.~\eqref{eq:EOMfermions} over $t$, leading to 
\EqS{
&\hat c_{\vk\a}^I(t)=i\hat c^0_{\vk\a}-\\
-&i\int_0^t dt'
\sum_{\vq,n} g_{n} e^{(\D \ve_{\vQ_n}-\w_0(\vq)) t'}\hat c^I_{\vk+\vQ_n,\a'}\hat b^I_\vq-\\
-&i\int_0^t dt'
\sum_{\vq,n} g_{n}e^{(\D \ve_{-\vQ_n}+\w_0(\vq)) t'}\hat c^I_{\vk-\vQ_n,\a'}(\hat b_\vq^I)\dg,
}
where $\D\ve_\vQ=\ve_{\vk+\vQ}-\ve_{\vk}$ and $\hat c^0_{\vk\a}=\hat c^I_{\vk\a}(0)$. To the linear order in $g_n$, we can approximate the operators $\hat c^I$ and $\hat b^I$ on the r.h.s. by $\hat c^0$ and $\hat b^0$, defined for $g_n=0$. Therefore, to this order in $g_n$, 
\EqS{
\hat c_{\vk\a}^I(t)=&i\hat c^0_{\vk\a}+
\sum_{\vq,n} g_{n} F(\D \ve_{\vQ_n}-\w_0(\vq),t)\hat c^0_{\vk+\vQ_n,\a'}\hat b^0_\vq+\\
+&
\sum_{\vq,n} g_{n}F(\D \ve_{-\vQ_n}+\w_0(\vq),t)\hat c^0_{\vk-\vQ_n,\a'}(\hat b_\vq^0)\dg,
}
where $F(\D\ve,t)=-i\int_0^t dt' e^{i \D\ve t'}=\frac{1-e^{i\D\ve t}}{\D\ve}$.
Next, we compute $\av{(\hat c^I_{\vk\a})\dg\hat c^I_{\vk\a}}$, yielding
\EqS{
f_{\vk\a}=&f^0_{\vk\a}+\sum_{\vq,n}|F(\D \ve_{\vQ_n}-\w_0(\vq),t)|^2 f^0_{\vk+\vQ_n,\a} n_{\vq}+\\
+&\sum_{\vq,n}|F(\D \ve_{-\vQ_n}+\w_0(\vq),t)|^2 f^0_{\vk-\vQ_n,\a} n_{\vq}.
}
where $f_{\vk\a}=\av{ (\hat c^I_{\vk\a})\dg\hat c^I_{\vk\a}}$, $f^0_{\vk\a}=\av{ (\hat c^0_{\vk\a})\dg\hat c^0_{\vk\a}}$ and $n_{\vq}=\av{(b_\vq^0)\dg b^0_\vq}\eqa \av{b^0_\vq(b_\vq^0)\dg }$. In the long time limit, $t\to \infty$, we approximate $|F(\D \ve,t)|^2\to 2\p t\dl (\D\ve)$. Differentiating over time, we arrive at
\EqS{
\dot f_{\vk\a}=&2\p \sum_{\vq,n}\dl(\D \ve_{\vQ_n}-\w_0(\vq)) f^0_{\vk+\vQ_n,\a} n_{\vq}+\\
+&2\p\sum_{\vq,n}\dl(\D \ve_{-\vQ_n}+\w_0(\vq)) f^0_{\vk-\vQ_n,\a} n_{\vq}.
}

\section{Analytical estimation of the gain}
Here, we analytically estimate the gain in the toy model [given in Eq.~(1) in the main text], deriving Eq.~(8) in the main text.  Our goal is to evaluate the period-averaged power, $\cP^0_\ye(\w)$, of the phonon mode with frequency $\w$, given by Eq.~(6) in the main text in the small-gain limit, $\g_\w\to 0$. For simplicity, we assume $\cM^{\a\a'}_{\vk,\vk'}=\frac{(2\p)^2}{\cA}q^2|\vu_0|^2g_1^2\av{\cO}^2\sum_{n=\pm}\dl(\vk-\vk'-\vQ_{n})$, where we defined $\vQ_{\pm}=(q\pm k_u)\hat\vx$. Therefore, the gain can be written as $\cP_\ye(\w)=\sum_{\a\a',n}[\cP^+_{\a\a',n}(\w)-\cP^-_{\a\a',n}(\w)]$, where 
\Eq{
\cP^\z_{\a\a',n}=2\p\cP_0\int \frac{d^2\vk}{(2\p)^2} f_{\a\vk} \dl(\ve_{\a}(\vk)-\ve_{\a'}(\vk-\z\vQ_n)+\z\w),
}
$\cP_0=\w q^2|\vu_0|^2g_1^2\av{\cO}^2$,  and $\z=\pm$.
We also assume $f_{\a\vk}=0$ for $0<\ve_{\a}(\vk)<V$, and $f_{\a\vk}=1$, otherwise. 
The expression breaks up into overlaps of two cones in the energy-momentum space, $\ve_{\a}(\vk)$ and $\ve_{\a'}(\vk-\vQ_\z)\pm\w$, where $\ve_\a(\vk)=\a v_\ye|\vk|$, for $\a=\pm$ [see below Eq.~(1) in the main text].

First, we evaluate $\cP^+_{++,n}$ corresponding to the gain in the system. It accounts for the intersection of $\ve_+(\vk)$ and $\ve_{+}(\vk-\vQ_n)+\w$ in the range $\max(V,\w)<\ve_+(\vk)<V+\w$. For $\w>\w_{Rn}$ [see Eq.~(3) in the main text for the definition of $\w_{R\pm}$], the two cones do not intersect, giving rise to zero contribution to the gain. In the opposite case, when $\w<\w_{Rn}$, the two cones intersect along a line. The maximal overlap is expected when $\w\lesssim\w_{Rn}$, where the two cones are nearly tangential. Focusing on this case, we define momentum in a spherical system of coordinates, such that $\vk=(k \cos(\q),k\sin(\q))$, where $k=|\vk|$ and $\q$ is the angle of $\vk$ relative to the $\hat\vx$ axis. The intersection of the cones occurs near $\q_1=0$ when $q+n k_u>0$ or near $\q_2 =\p$ when $q+n k_u<0$. 
We expand $\ve_{+}(\vk-\vQ_n)$, to the leading (quadratic) order in $\dl\q=\q-\q_{1/2}$, yielding
$\ve_+(\vk-\vQ_\z)=v_\ye|k-Q_n|+\frac{v_\ye k Q_n}{2|k-Q_n|}\dl\q^2+\cO(\dl\q^4)$, where $Q_n=|\vQ_n|$. At this order in $\dl\q$, the intersection $\ve_+(\vk)=\ve_+(\vk-\vQ_n)+\w$, occurs at $\dl\q_0^{\pm}=\pm \sqrt{(v_\ye Q_n-\w)\frac{2(k-Q_n)}{v_\ye k Q_n}}$, where we considered $k> Q_n$, corresponding to the dominant contribution. Note, that $v_\ye Q_n=\w_{Rn}$ identically.

Next, we split the integral in Eq. (1) to an integral over the energy $\ve=v_\ye k$ and the angle $\dl\q$, as $\int \frac{d^2\vk}{(2\p)^2}=\int d\ve \cN_D(\ve)\int_0^{2\p} \frac{d\q}{2\p}$. The density of states for the dispersion $\ve_+(\vk)$ is given by $\cN_D(\ve)=\ve/(2\p v_\ye^2)$. The energy conserving $\dl$-function can be simplified in the angular coordinates as 
$\dl(\ve_+(\vk)-\ve_+(\vk-\vQ_n)-\w)=\frac{\ve-\w_{Rn}}{ \ve \w_{Rn}|\dl\q_0^+|}\sum_{i=\pm}\dl(\dl\q-\dl\q^i_0)$. The angular integral therefore can be trivially performed due to the $\dl$-function, leaving the energy integral
\Eq{
\cP_{++,n}^+=\cP_0\int_{\max(\w,V)}^{\w+V} d\ve \frac{2(\ve-\w_{Rn})}{ \ve \w_{Rn}|\q_0^+|}\cN_D(\ve).
}
 We implicitly assume that $\cP_{++,n}^+=0$ for $\w>\w_{Rn}$ or $\ve<\w_{Rn}$. We therefore obtain
\Eq{
\cP^+_{++,n}=\frac{\cP_0\cI^+_n(\w,V)\Q(\w_{Rn}-\w)}{\p v_\ye^2\sqrt{2\w_{Rn}(\w_{Rn}-\w)}},
}
where $\cI^+_n(\w,V)=\int_{\max(\w_{Rn},\w,V)}^{\max(\w_{Rn},\w+V)}d\ve \sqrt{\ve}\sqrt{\ve-\w_{Rn}}$, which has an exact analytic expression. In the limit $V\gg\w_{Rn}$, $\cI^+_n(\w,V)=V\w_{Rn}$.

Next, we evaluate $\cP^-_{-+,n}$ accounting for an intersection of $\ve_{-}(\vk)$ and $\ve_+(\vk+\vQ_n)-\w$ in the range $-\w<\ve_+(\vk)<\min(0,V-\w)$ contributing to a negative gain (i.e., absorption of the phonons by the electrons). The intersection occurs at $\dl\q_0^\pm=\pm\sqrt{(\w-v_\ye Q_n)\frac{2(Q_n-k)}{v_\ye k Q_n}}$ for $k<Q_n$. Integration over the energy and angle in the range $\ve\in[\max(0,\w-V),\w]$, using $\dl(\ve_-(\vk)-\ve_+(\vk+\vQ_n)+\w)=\frac{\w_{Rn}-\ve}{ \ve \w_{Rn}|\dl\q_0^+|}\sum_{i=\pm}\dl(\dl\q-\dl\q^i_0)$, yields
\Eq{
\cP^-_{-+,n}=\frac{\cP_0\cI^-_n(\w,V)\Q(\w-\w_{Rn})}{\p v_\ye^2\sqrt{2\w_{Rn}(\w-\w_{Rn})}},
}
where $\cI^-_n(\w,V)=\int_{\min(\w_{Rn},\max(0,\w-V))}^{\w_{Rn}}d\ve \sqrt{\ve}\sqrt{\w_{Rn}-\ve}$. 
The total contribution is shown in Fig.~2b in the main text. For $\w=\w_{Rn}-\dl \w_n$, where $\dl\w_n\ll\w_{Rn}$ and in the limit $V\gg\w_{Rn}$, we approximate $\cP_\ye(\w)\eqa \cP_0 \cN_D(V)\sqrt{\frac{ 2\w_{Rn}}{\dl\w_n}} $.
Therefore, the gain, as follows from Eq.~(7) in the main text reads
$\g_\w=\frac{\w g_1^2\av{\cO}^2 a^2}{c_{\rm ph}^3M}\cN_D(V)\sqrt{\frac{ 2\w_{Rn}}{\dl\w_n}}$.

\section{The Bistrizer-MacDonald model of the TBG}
In this section, we present the continuum model that describes the low-energy physics of the TBG near the charge neutrality. Our goal is to outline the model that we used in the numerical simulations and to estimate the energy scale of the spatially modulated electron-phonon coupling due to the uniaxial strain and an array of screening gates. Therefore, we begin with a continuum model with a generic weak strain and screened electron-phonon interaction.

The TBG consists of two graphene monolayers twisted by a relative
angle $\q$. Each untwisted graphene monolayer, denoted by $l=1,2$, comprises  a honeycomb lattice of carbon atoms with a lattice constant of $a=0.246\, {\rm nm}$ and reciprocal vectors  $\vG_{1,2}=\frac{2\p}{a}(1,\pm \frac{1}{\sqrt{3}})$.
The graphene band structure exhibits two valley points at $\vK_{\pm}=\mp (\vG_1+\vG_2)/3$, where the electrons have nearly Dirac dispersion. Our model is linearized around the valley points giving rise to the valley number, $\x=\pm$.
Under a uniaxial strain and a twist, the low energy physics of each monolayer, at valley $\x$, is given by \cite{Bi2019,Koshino2020} 
\Eq{
h^{(l)}_{\x}=- v_F [(\I+\cE^T_l)(\vk-\vcD^{(l)}_{\x}+\x\vcA^{(l)}_{\rm ph})]\cdot(\x \s^x,\s^y)+\F^{(l)}_{\rm ph}.
\label{eq:MonolayerHamiltonian}
}
Here, $v_F/a=2.14\, {\rm eV}$, $\s^x$ and $\s^y$ are Pauli matrices acting in the sublattice basis $\bC{\ket{A},\ket{B}}$ of each monolayer ($\s^x\ket{A}=\ket{B}$), $\cE_l$ is the strain and rotation tensor, given for small deformations by $\cE_l= \mat{\e_{xx}&\e_{xy}-\q_l\\ \e_{yx}+\q_l&\e_{yy}}$, where $\q_{1,2}=\pm \q/2$ is the rotation angle of each monolayer.
The vector $\vcD^{(l)}_{\x}=(\I-\cE_l^T)\vK_\x-\x\vcA$ includes the modified position of the valley points due to the deformation and an effective gauge connection imposed by the strain, reading $\vcA=\frac{\sqrt{3}}{2a}\be (\e_{xx}-\e_{yy},\e_{xy})$, with $\be\eqa 3.14$.  Eq.~\eqref{eq:MonolayerHamiltonian} also includes coupling to acoustic phonons, represented by the effective gauge connection $\vcA^{(l)}_{\rm ph}=\frac{\sqrt{3}}{2a}\be (\hat u^{(l)}_{xx}-\hat u^{(l)}_{yy},\hat u^{(l)}_{xy})$ and by the diagonal term $\F_{\rm ph}^{(l)}=D (u^{(l)}_{xx}+u_{yy}^{(l)})$. Here, $\hat u_{ij}=(\dpa_i \hat u_j^{(l)}+\dpa_j \hat u_i^{(l)})/2$, and $\hat u_i^{(l)}(\vr,t)$ is the displacement operator in the direction $i=\{x,y\}$ and layer $l$. Each monolayer has an additional spin degree of freedom which is degenerate in the model.

For $\e_{ij}=0$ and $\be=D=0$, the twisted structure exhibits an emergent moir\'e lattice with a honeycomb structure, described by the reciprocal vectors $\vG^\ym_{1,2}=G_\ym(\pm \frac{1}{\sqrt{3}},1)$, where $G_\ym=\frac{2\p}{a_\ym}$ and  $a_\ym=a/[2 \sin(\q/2)]$. 
The Hamiltonian describing the low-energy physics of the TBG is obtained by combining Eq.~\eqref{eq:MonolayerHamiltonian} describing monolayers and interlayer hopping terms $T_\x$, yielding\cite{Bistritzer2011,Koshino2019,Bi2019} for the valley $\x$,
\Eq{
\cH_\x=\mat{h^{(1)}_{\x}&T\dg_\x\\T_\x&h^{(2)}_{\x}}.
\label{eq:TBGHamiltonian}
}
Here, the interlayer hopping is approximated by $T_\x(\vr)=\mat{u&u'\\u'&u}+\mat{u&u'w\inv\\u'/w&u}e^{i\x\vG^\ym_1\cdot \vr}+\mat{u&u'w\\u'/w\inv&u}e^{i\x\vG^\ym_2\cdot \vr}$ in the sublattice basis, where $w=e^{i2\p\x/3}$, and we consider $u=0.0797\, {\rm eV}$, $u'=0.0975\, {\rm eV}$.
Fig.~3a in the main text shows the spectrum of a single valley of the TBG, described by Eq.~\eqref{eq:TBGHamiltonian}, for $\e_{ij}=\be=D=0$.

Next, we discuss the realizations of the nano-undulator. We begin by discussing a spatially modulated uniaxial strain along the $x$ direction. We parametrize such a strain by $\e_{xx}=\e_0 \cos(k_u x)$, where $\e_0$ is the amplitude of the strain, assumed to be small $\e_0\ll1$, and $\e_{xy}=\e_{yx}=\e_{yy}=0$. The effective electron-phonon coupling in the presence of the strain is given by the linear in $\e_0\hat u^{(l)}_{ij}$ term in Eq.~\eqref{eq:MonolayerHamiltonian} . Expanding to this order we arrive at 
\Eq{
h^{\rm e-p}_{l}=- v_F \frac{\sqrt{3}}{2a}\be\e_{0}\cos(k_u x)  (\hat u_{xx}^{(l)}-\hat u_{yy}^{(l)})\s^x.
\label{eq:ElectronPhonon}
}
By comparing Eq.~\eqref{eq:ElectronPhonon} and Eq.~(2) in the main text, we find $g_1=\frac{\sqrt{3}}{4a}v_F \be\e_{0}$.
For $\e_0\eqa 5\%$ strain, we estimate $g_1\eqa 0.15\, {\rm eV}$, corresponding to $\g_0\eqa 0.02\, \rm \m m\inv$.

Another realization is based on a periodic array of gates placed at the distance $d$ from the TBG. The gates modify the screening efficiency of the interaction between the electronic charge density and the lattice ions\cite{Mariani2010,Hwang2007,VonOppen2009,Suzuura2002a,Manes2007,Sohier2014,Park2014}. For a phonon of momentum $q$, the renormalized coupling term near the gate can be estimated by $D=D_0 q/[q+q_{\rm TF}\tanh(q d)]$, where we use the estimate for the Thomas Fermi wavevector\cite{Kim2020} $q_{\rm TF}\eqa 1\, {\rm nm}\inv$ . For a periodic structure of gates along $\hat\vx$ with periodicity $\lm_u$, and in the limit $q_{\rm TF}\gg q, 1/d$, the coupling approximately oscillates between $D_{\rm min}\eqa \frac{D_0q}{q_{\rm TF}}$ and $D_{\rm max}\eqa \frac{D_0}{1+q_{\rm TF}d}$.  
Comparing with Eq.~(2) in the main text, we estimate $g_1\eqa \frac{1}{4}\frac{D_0}{1+q_{\rm TF }d}$ and $\cO\sim \I$. For $q_{\rm TF} d\eqa 3$ and $D_0= 50\, {\rm eV}$, we estimate $g_1\eqa 3\, {\rm eV}$. For this value of $g_1$, we obtain $\g_0\eqa 8.5\, {\rm \m m }\inv$.

\section{Coupling of the phonons to plasmon modes}
In this section, we discuss the coupling of phonons to plasmon modes, giving rise to an evanescent THz-oscillating electromagnetic field. To describe this effect, we assume a uniform density of electrons $\ro^0_\ye$ modulated by small fluctuations represented by the displacement operator $\vu_\ye(\vr,t)$, resulting in the density operator, $\hat\ro_\ye(\vr,t)=\ro_0[1-\nabla\cdot \hat\vu_\ye(\vr,t)]$. Similarly, the density operator of the ions modulated by the phonon displacement field $\hat\vu(\vr,t)$ [see the main text], is given by $\hat\ro_L(\vr,t)=\ro^0_L[1-\nabla\cdot \hat\vu(\vr,t)]$, where $\ro^0_L$ is the density of ions. For uniform densities (when $\hat\vu_\ye=\hat\vu=0$), the sample is neutral, corresponding to $\ro_0=Z_L\ro^0_L$, where $Z_L$ is the ions' charge. Plasma modes in the electronic density and phonon modes in the ions, give rise to local electric charge density given by $\hat\dl\ro(\vr,t)=\ro_0\nabla\cdot [\hat\vu(\vr,t)-\hat\vu_\ye(\vr,t)]$. This charge density, in turn, creates an oscillating electric field.

The dynamics of the electronic density fluctuations is described by the Hamiltonian 
\Eq{
\hat\cH_\ye=\int \frac{d^2\vr}{a^2_\ym} \bS{\frac{1}{2m_\ye}|\hat\vpi_\ye(\vr)|^2+\frac{\ka_\ye}{2}|\dpa_\vr\hat \vu_\ye(\vr)|^2}.
}
Here, $\hat \vpi_\ye(\vr)$ is the conjugate momentum of $\hat \vu_\ye(\ve)$, $m_\ye$ is an effective electron mass\cite{Svintsov2012} and $\ka_\ye$ is related to the electronic compressibility, which near the Dirac cone [see Eq.~(1) in the main text], can be approximated as $\ka_\ye\eqa  v_\ye/a_\ym$, where $a_\ym=a/(2\sin(\q/2))$.
Similarly, the lattice displacement is described by the Hamiltonian 
\Eq{
\hat\cH_L=\int \frac{d^2\vr}{a^2} \bS{\frac{1}{2M}|\hat\vpi(\vr)|^2+\frac{M c^2_{\rm ph}}{2}|\dpa_\vr\hat \vu(\vr)|^2}.
}
The electronic and the ionic densities are coupled by the Coulomb repulsion between the charge densities $e\hat\dl \ro$ at different positions, which is described by the Hamiltonian
\Eq{
\hat\cH_{C}=\iint d^2\vr d^2\vr' V_\yC(\vr-\vr') \dl\hat \ro(\vr)\dl \hat\ro(\vr'),
}
where $V_\yC(\vr-\vr')=e^2/|\vr-\vr'|$. We also consider the electron-phonon coupling [see Eq.~(2) in the main text] written in the form
\Eq{
\hat\cH_{\rm ep}=-\int d^2\vr g(\vr)\ro_\ye^0 (\nabla\cdot \hat\vu_\ye) (\nabla\cdot \hat \vu).
}
The equation of motion for the displacement operators in the Heisenberg picture driven by the Hamiltonian $\hat\cH=\hat\cH_{\ye}+\hat\cH_L+\hat\cH_C+\hat\cH_{\rm ep}$, reads
$\dpa_t\hat\vu_\ye=\hat\vpi_\ye/m_\ye$ and $\dpa_t\hat\vu=\hat\vpi/M$. In turn, the equation of motion for the conjugate momenta reads
\Eq{
\dpa_t\hat\vpi_\ye=\ka_\ye \dpa^2_\vr \hat\vu_\ye-2a_\ym^2\dpa_\vr \hat\f(\vr)-a_\ym^2 g(\vr)\ro_0 \dpa_\vr^2 \hat\vu,
}
where $\hat\f(\vr)=\int d^2\vr' V_{\yC}(\vr-\vr')\ro_\ye^0 \dl\hat \ro(\vr')$.
Similarly, for the phonon conjugate momentum, we obtain 
\Eq{
\dpa_t\hat\vpi=M c^2_{\rm ph} \dpa^2_\vr \hat\vu_\ye+2a^2\dpa_\vr \hat\f(\vr)-a^2 g(\vr)\ro_0 \dpa_\vr^2 \hat\vu_\ye.
}
Differentiating over time the equations of motion of the displacement fields and combining with equations of motion for the conjugate momenta, we arrive at 
\EqL{
&&M\dpa^2_t \hat\vu=M c^2_{\rm ph} \dpa^2_\vr \hat\vu+2a^2\dpa_\vr \hat\f(\vr)-a^2g(\vr)\ro_0 \dpa_\vr^2 \hat\vu_\ye\\
&&m_\ye\dpa^2_t \hat\vu_\ye=\ka_\ye \dpa^2_\vr \hat\vu_\ye-2a_\ym^2\dpa_\vr \hat\f(\vr)-a_\ym^2 g(\vr)\ro_0 \dpa_\vr^2 \hat\vu.
}
To find the eigenmodes of the coupled differential equation, we substitute $\hat \vu(\vr,t)=\frac{1}{\sqrt{\cA}}\hat \vu(\vq,\w) e^{i\vq \cdot \vr-i\w t}$ and $\hat \vu_\ye(\vr,t)=\frac{1}{\sqrt{\cA}}\hat \vu_\ye(\vq,\w) e^{i\vq \cdot \vr-i\w t}$, leading to 
\EqL{
&&\w^2 \hat\vu= c^2_{\rm ph} q^2 \hat\vu+\frac{aqv_\yC}{M}(\hat\vu-\hat\vu_\ye)- \frac{a^2q^2v_\yD}{M} \hat\vu_\ye\\
&&\w^2 \hat\vu_\ye=\frac{\ka_\ye q^2}{m_\ye} \hat\vu_\ye-\frac{a^2_\ym q v_\yC}{a m_\ye}(\hat\vu-\hat\vu_\ye)-\frac{a_\ym^2 q^2 v_\yD}{m_\ye}\hat\vu,
}
where we used the surface Fourier transform of $\cF\bC{V_{\yC}}=2\p e^2 /q$, take only the constant in space component of $g(\vr)$, and defined $v_\yC=4\p e^2 \ro_0^2 a$, $v_\yD=g_0\ro_0$. 
We can rewrite the latter equation as an eigenvalue problem
\Eq{
\w^2 \vec U=\cK \vec U,
}
where $\cK=\mat{c_{\rm ph}^2 q^2+\frac{a q v_\yC }{M}&-\frac{a_m q v_\yC}{\overline{M}}-\frac{a_\ym aq^2v_\yD}{\overline{M}}\\-\frac{a_\ym q v_\yC}{\overline{M}}-\frac{a_m a q^2v_\yD}{\overline{M}}&
\frac{\ka_\ye}{m_\ye}q^2+\frac{a_\ym^2 q v_\yC}{a m_\ye}}$, $\vec U=((a_\ym/a)\hat \vu/\sqrt{m_\ye},\hat\vu_\ye/\sqrt{M})^T$, and $\overline{M}=\sqrt{M m_\ye}$. 
The eigenvalues to the leading order in $q$ read 
\EqL{
&&\w^2_+= \frac{a_\ym^2 q v_\yC}{a m_\ye}\bR{1+\frac{a^2m_\ye}{a_\ym^2 M}}+\cO(q^2)\\
&&\w^2_-= \frac{q^2}{1+\frac{a^2m_\ye}{a_\ym^2M}}\bR{c_{\rm ph}^2+\frac{a^2\ka_\ye}{a_\ym^2 M}-2a^2v_\yD}+\cO(q^3),
}
respectively corresponding to the plasmon and phonon modes. The plasmon eigenmode, corresponding to $\w_+$, reads  $\hat\vu_\ye/\hat\vu=-\frac{a_\ym M}{a m_\ye}+\cO(q)$.  
Similarly, the phonon eigenmode, corresponding to $\w_-$, reads $\hat\vu_\ye/\hat\vu=1+\lm q+\cO(q^2)$, where
\Eq{
\lm=a\frac{(a_\ym^2 M-a^2 m_\ye)v_\yD}{(a_\ym^2 M+a^2 m_\ye)v_\yC}+a\frac{c^2_{\rm ph}m_\ye M-M\ka_\ye}{(a_\ym^2 M+a^2 m_\ye)v_\yC}.
\label{eq:Lambda}
}
Therefore, for $q\to 0$ phonons, $\hat\vu_\ye\to\hat\vu$, corresponding to $\dl\ro\to0$.
For a finite momentum excitation, we find \Eq{
\dl\hat\ro(\vr,t)=\ro_0 \lm q^2 \hat\vu(\vr,t).
}

For $\vq=q\hat\vx$ and $\w=\w_-$, the charge density wave in the $\hat \vx$ direction gives rise to an oscillating current density in the same direction, as follows from the continuity equation $J_x(\vr,t)=-\frac{\w}{q}\av{\dl\hat\ro(\vr,t)}$. The current density generates an oscillating magnetic field which near the sample plane ($z=0$) is oriented mostly in the $\hat \vy$ direction and can be estimated by Ampere's law as $B_y(\vr,t)=\frac{2\p e}{c_\ell}J_x(\vr,t)$, where $\vr=(x,y,z=0)$ and $c_\ell$ is the speed of light. For $z>0$, this oscillating magnetic field propagates according to the electro-magnetic wave equation. Assuming $B_y(\vr,t)=B_y(z=0,t)e^{i k_z z}$, we find $k_z^2=\w^2/c_\ell^2-q^2$. Since, $c_\ell\gg c_{\rm ph}$, $k_z$ obtains imaginary values, corresponding to an evanescent electromagnetic wave. The corresponding electric field is given by $\vec E=-i\frac{c_\ell}{\w}\vec \nabla\times\vec B$, yielding
$\vec E(\vr,t)=-\frac{c_\ell}{\w}(\hat \vx k_z-\hat \vz q)B_y(\vr,t)$.
In terms of the charge densities, the electric field reads
\EqL{
&&E_x=2\p e\av{\dl \hat\ro}  e^{-|k_z| z}(k_z/q)\\
&&E_z=-2\p e \av{\dl \hat\ro} e^{-|k_z| z},
}
corresponding to the amplitude
\Eq{
|\vec E|=2\p e \ro_0 \lm q^2 |\av{\hat\vu(\vr,t)}| e^{-|k_z| z}\sqrt{1+|k_z/q|^2}.
}

We estimate Eq.~\eqref{eq:Lambda} by $\lm\eqa a v_\yD/v_\yC$. Taking $\ro_0= 1/a_\ym^2$, and $g_0\eqa 50\, \rm eV$, we find $\lm\eqa 2 a$. Assuming at the laser saturation the displacement vector is  $|\av{\hat\vu}|\eqa 0.1 a$, taking $q a\eqa 10^{-2}$, and focusing on the near field $z\ll|k_z|\inv$, we estimate 
$|\vec E|\eqa 30\, \rm kV/m$.

\bibliography{Bibliography}